\documentclass[
               aps,                  
               prd,                  
               reprint,              
               nofootinbib,          
               superscriptaddress,   
               showpacs,             
               showkeys,             
               longbibliography      
              ]{revtex4-1}

\usepackage[utf8]{inputenc}
\usepackage{amsfonts,amsmath,amssymb}
\usepackage{bm,dsfont} 
\usepackage{url}
\usepackage{graphicx}
\usepackage{hyperref}

\newcommand{\nwse}[3]{\ensuremath{#1^{#2}_{\phantom{#2} #3}}}

\newcommand{\um}{\mathds{1}}

\newcommand{\udec}{Departamento de Física, Universidad de Concepción, Casilla 160-C, Concepción, Chile}
\newcommand{\unal}{Departamento de Física, Universidad Nacional de Colombia, Bogotá, Colombia}
\newcommand{\asctp}{Arnold Sommerfeld Center for Theoretical Physics, Ludwig-Maximilians-Universität München.\ Theresienstraße 37, 80333 München, Germany}
\newcommand{\uach}{Centro de Docencia de Ciencias Básicas para Ingeniería, Universidad Austral de Chile, Casilla 567, Valdivia, Chile}

\begin{document}

\title{Randall--Sundrum brane universe as a ground state for Chern--Simons gravity}

\author{Fabrizio Cordonier-Tello}
\email{f.cordonier@physik.uni-muenchen.de}
\affiliation{\asctp}

\author{Fernando Izaurieta}
\email{fizaurie@udec.cl}
\affiliation{\udec}

\author{Patricio Mella}
\email{patricio.mella@uach.cl}
\affiliation{\uach}

\author{Eduardo Rodríguez}
\email{eduarodriguezsal@unal.edu.co}
\affiliation{\unal}

\date{\today}


\begin{abstract}
In stark contrast with the three-dimensional case, higher-dimensional Chern--Simons theories can have non-topological, propagating degrees of freedom.
Finding those vacua that allow for the propagation of linear perturbations, however, proves to be surprisingly challenging.
The simplest solutions are somehow ``hyper-stable,'' preventing the construction of realistic, four-dimensional physical models.

Here, we show that a Randall--Sundrum brane universe can be regarded as a vacuum solution
of Chern--Simons gravity in five-dimensional spacetime, with non vanishing torsion along the dimension
perpendicular to the brane. Linearized perturbations around this solution not only exist, but behave as standard gravitational waves
on a four-dimensional Minkowski background. In the non-perturbative regime, the solution leads to a four-dimensional 
``cosmological function'' $\Lambda \left( x \right)$ which depends on the Euler density of the brane.

Interestingly, the fact that the solution admits nontrivial linear perturbations seems to be related to an often neglected property
of the Randall--Sundrum spacetime: that it is a group manifold, or, more precisely, two identical group manifolds glued together along the brane.
The gravitational theory is then built around this fact, adding the Lorentz generators and one scalar generator needed to close the algebra.
In this way, a conjecture emerges: a spacetime that is also a group manifold can be regarded as the ground state of a Chern--Simons theory
for an appropriate Lie algebra.
\end{abstract}



\maketitle

\section{Introduction}

Higher-dimensional Chern--Simons (CS) theories enjoy a host of properties that make them interesting theoretical laboratories.
For a recent review of their relevance to gravitation theory, the reader is advised to check Ref.~\cite{Zan15}.

We would like to focus here on a generic problem that appears in five and higher dimensions, namely,
that the simplest solutions are ``hyper-stable,'' meaning that no linearized perturbations can propagate on them.
In particular, we want to showcase how this problem can be overcome in a particular setting, and what we can learn from this.

The setting is a five-dimensional CS theory built upon the Weyl subalgebra of the conformal algebra,
which forgoes special conformal transformations to include only rotations, translations, and dilations.
This is a theory of gravity (in its first-order guise, as usual when dealing with CS theories) non-minimally coupled to a one-form field.
The details of the theory are given in section~\ref{sec:CStheory}, where the Lagrangian, gauge transformations,
and field equations are derived from the general properties of CS theories.
It must be noted, however, that the gravitational Lagrangian is not
Einstein--Hilbert's, but the dimensional continuation of the Euler density.
This term is a total derivative in four dimensions, but not in five,
where it leads to nontrivial dynamics~\cite{Lov71}.
In section~\ref{sec:dynamics}, we show that a Randall--Sundrum (RS)
brane universe~\cite{Ran99a,Ran99b},
with nonzero torsion along the direction perpendicular to the brane,
solves the CS field equations and admits linearized perturbations that propagate
as standard gravitational waves on a four-dimensional Minkowski background.

Why should this be so?
An attempt at answering this question begins in section~\ref{sec:RSBC}, where we show that the RS
brane universe can be regarded as two copies of half a certain group manifold, with the brane acting as a mirror between the two.
The commutation relations for the associated Lie algebra are explicitly displayed.
A CS theory built upon this algebra would not be a theory of gravity, however,
since there is no Lorentz symmetry and hence no spin connection or curvature.
As the most economical option, we embed the algebra in the Weyl subalgebra of the conformal algebra,
which includes the original generators, plus the Lorentz symmetry and a single extra generator.
It then comes as less of a surprise that the RS brane universe
is indeed a solution of a CS theory based upon the Weyl algebra.

While the resulting system may not quite turn out to be particularly realistic
(we discuss some of its cosmological implications in section~\ref{sec:dynamics},
where we show that it can lead to unphysical consequences),
and should therefore be regarded as a toy model,
we believe that its main interest lies in the lesson that is suggests:
to construct an admissible ground state for a CS theory, look for a solution that admits a group structure.

Our method can be briefly described as follows.

Let $\bm{A}$ be a Lie algebra-valued one-form gauge connection
and let $\bm{F} = \mathrm{d} \bm{A} + \frac{1}{2} \left[ \bm{A}, \bm{A} \right]$
be its associated gauge curvature.
In $d=2n+1$ dimensions, the CS field equations can be written as
$\left\langle \bm{F}^{n} \bm{G}_{A} \right\rangle = 0$,
where the $\bm{G}_{A}$ generators span a basis for the gauge algebra,
and $\left\langle \cdots \right\rangle$ stands for a
multilinear symmetric form of rank $n+1$ invariant under the algebra,
such as the symmetrized trace in a suitable matrix representation.
The simplest solutions satisfy $\bm{F} = 0$, which,
for $n \geq 2$, means that they are hyper-stable:
no linear perturbations propagate around this background.%
\footnote{This is sometimes referred to as the ``linearization instability''
problem in the literature.}
This happens because perturbations must satisfy
$\left\langle \bm{F}^{n-1} \delta \bm{F} \bm{G}_{A} \right\rangle = 0$,
which becomes empty when $\bm{F} = 0$.
For $n=1$, on the other hand, perturbations are not affected by the fact that $\bm{F} = 0$,
and indeed there exist interesting cases of propagating fields in three-dimensional CS theory
(see, e.g., Ref.~\cite{Car01}).
This is even more remarkable given the fact that, provided the algebra is semisimple
(so that the bilinear form $\left\langle \cdots \right\rangle$ is invertible),
the CS field equations in three dimensions are actually equivalent to $\bm{F} = 0$.
A higher-dimensional vacuum that admits linear perturbations must necessarily
have a non-vanishing gauge curvature, all the while satisfying
$\left\langle \bm{F}^{n} \bm{G}_{A} \right\rangle = 0$.

Pick now a spacetime that is also a group manifold,
meaning that its vielbein $E^{A}$ can be regarded
as the Maurer--Cartan (MC) forms, $\bm{\theta} = \frac{1}{l} E^{A} \bm{Z}_{A}$,
of a certain Lie algebra spanned by the $\bm{Z}_{A}$ generators.
These MC forms satisfy
$\mathrm{d} \bm{\theta} + \frac{1}{2} \left[ \bm{\theta}, \bm{\theta} \right] = 0$.
If $\bm{A}$ is given by
$\bm{A} = \bm{\theta} + \bm{W} + \cdots$,
where $\bm{W}$ is the Lorentz algebra-valued spin connection,
and the dots stand for any additional fields, then the gauge curvature reads
\begin{equation}
 \bm{F} = \mathrm{d} \bm{W} + \frac{1}{2} \left[ \bm{W}, \bm{W} \right] +
 \mathrm{d} \bm{\theta} + \frac{1}{2} \left[ \bm{\theta}, \bm{\theta} \right] +
 \left[ \bm{W}, \bm{\theta} \right] + \cdots .
\end{equation}
Here, we identify
$\bm{R} = \mathrm{d} \bm{W} + \frac{1}{2} \left[ \bm{W}, \bm{W} \right]$ and
$\bm{T} = \mathrm{d} \bm{\theta} + \left[ \bm{W}, \bm{\theta} \right]$ with
the Lorentz curvature and torsion, respectively.

The fact that spacetime is a group manifold now has several consequences.
First of all, since
$\mathrm{d} \bm{\theta} + \frac{1}{2} \left[ \bm{\theta}, \bm{\theta} \right] = 0$,
some components of the gauge curvature vanish.
On the other hand, there's no a priori reason why torsion should vanish,
so these models in general feature nonzero torsion.
The torsional degrees of freedom can be parameterized as
$T^{A} = \nwse{\kappa}{A}{B} E^{B}$, where $\kappa^{AB}$
is the cotorsion one-form.
The cotorsion then determines the spin connection via
$W^{AB} = \mathring{W}^{AB} + \kappa^{AB}$,
where the torsionless component $\mathring{W}^{AB}$
is found by solving the equation
$\mathrm{d} E^{A} + \nwse{\mathring{W}}{A}{B} E^{B} = 0$.
In turn, the torsionless part of the spin connection is used to define
the Riemann curvature,
$\nwse{\mathring{R}}{A}{B} = \mathrm{d} \nwse{\mathring{W}}{A}{B} +
 \nwse{\mathring{W}}{A}{C} \nwse{\mathring{W}}{C}{B}$.
Finally, the Lorentz curvature is computed as
$\nwse{R}{A}{B} = \nwse{\mathring{R}}{A}{B} +
 \mathring{\mathrm{D}} \nwse{\kappa}{A}{B} +
 \nwse{\kappa}{A}{C} \nwse{\kappa}{C}{B}$.

To summarize, the metric structure encoded in the vielbein
determines the torsionless component of the spin connection
and the Riemann curvature, while the cotorsion is needed
to compute the full Lorentz curvature.
The fact that spacetime has a group structure means that
the problem of finding a general background that allows for
linear perturbations to propagate has been reduced to finding
a one-form cotorsion with this property, which proves to be more tractable.

We close with some final thoughts and an outlook for future work in section~\ref{sec:final}.

\section{Randall--Sundrum and the Bianchi Classification}
\label{sec:RSBC}

Although far from self-evident, the RS metric is intrinsically related to the Weyl subgroup of the conformal group in any dimension.
The easiest way to see this is by means of the MC approach to Lie groups~\cite{deAz95}.

Let $M$ be a spacetime manifold endowed with a notion of metricity $g$ codified through a basis of vielbein one-forms
$E^{A} = \nwse{E}{A}{\mu} \mathrm{d} x^{\mu}$,
\begin{equation}
 g = \eta_{AB} E^{A} \otimes E^{B}
   = g_{\mu\nu} \mathrm{d} x^{\mu} \otimes \mathrm{d}x^{\nu}
   \label{Eq_Metric}.
\end{equation}
Consider now a constant vector $k^{A}$ and a basis $m^{A} = \nwse{m}{A}{\mu} \mathrm{d} x^{\mu}$ of vielbein one-forms for the Minkowski spacetime.
Then it is straightforward to show that, when
\begin{equation}
 E^{A} \left( x \right) = - \frac{1}{k_{B} \nwse{m}{B}{\lambda} x^{\lambda}} m^{A},
 \label{Eq_MC_Vielbeine}
\end{equation}
then $M$ is the higher-dimensional analogue of a three-dimensional Type~V Bianchi spatial section~\cite{Landau_Lifshitz}.

As matter of fact, by defining the dimensionless one-form
\begin{equation}
 \theta^{A}=\frac{1}{l}E^{A},
\end{equation}
where $l$ is a constant parameter with units of length,%
\footnote{We prefer the convention where the structure constants of a Lie algebra are always dimensionless,
and therefore the MC one-forms $\theta^{A}$ must be dimensionless too.
Since the metric $g=g_{\mu\nu}\mathrm{d}x^{\mu}\otimes\mathrm{d}x^{\nu}$
has units of length squared, the vielbein one-form $E^{A}$ has units of length.
That is why it is necessary to introduce a constant parameter $l$ with units of length in order to make everything dimensionally consistent.
The same has to be done in eq.~(\ref{Eq_Weyl_Connection}).
From a physical point of view it may seem natural to identify $l$ with the Planck length. The presence of $l$ can partially
determine the structure of a theory; see, e.g., Ref.~\cite{Iza11a}.}
one can prove that $\theta^{A}$ satisfies
\begin{equation}
 \mathrm{d}\theta^{C}=-\frac{1}{2}l\left(\delta^{C}_{A}k_{B}-\delta^{C}_{B}k_{A}\right)\theta^{A}\wedge\theta^{B},
\end{equation}
and can therefore be regarded as the MC one-form for the associated Lie algebra
\begin{equation}
 \left[\bm{Z}_{A}, \bm{Z}_{B}\right] = l \left( \delta^{C}_{A} k_{B} - \delta^{C}_{B} k_{A} \right) \bm{Z}_{C}.
 \label{eq:BTValg}
\end{equation}
Eq.~(\ref{eq:BTValg}) is of course the generalization to an arbitrary dimension of the Bianchi Type~V algebra,
implying that $M$ is in this case also a group manifold.
This is not strange; the Bianchi classification of three-dimensional spaces has been very well studied
because of its importance in cosmology, but the general concept can be defined in any dimension.

It is straightforward to check that the vielbein~(\ref{Eq_MC_Vielbeine}) induces a warped metric $g=\eta_{AB}E^{A}\otimes E^{B}$.
For clarity, let us choose a five-dimensional spacetime and a range of indices $A, B, \ldots = 0, 1, \ldots, 4$.
When $k^{A}$ is spacelike, it possible to rotate the f\"{u}nfbein in such a way that the only non-vanishing component of $k_{A}$ is $k_{4}=k$.
After making the standard transformation of coordinates $x^{4}=e^{kz}/k$, the associated metric becomes
\begin{equation}
 g=e^{-2kz}\eta_{\mu\nu}\mathrm{d}x^{\mu}\otimes\mathrm{d}x^{\nu}+\mathrm{d}z\otimes\mathrm{d}z.
\end{equation}

This means that the RS metric with a flat brane,
\begin{equation}
 \bar{g} = e^{-2k \left\vert z \right\vert}
 \eta_{\mu\nu} \mathrm{d} x^{\mu} \otimes \mathrm{d} x^{\nu} +
 \mathrm{d} z \otimes \mathrm{d} z,
\end{equation}
may be regarded as describing half the group manifold, glued with its reflection on the $z=0$ brane.
The direction orthogonal to the brane is that of the vector $k^A$.

Our purpose is to construct an off-shell invariant theory of gravity in $d=5$
that admits the RS metric as a solution, with well-defined perturbations around it.
To achieve this goal, one may try a CS theory for the simple one-form gauge connection
\begin{equation}
 \bm{A}=\frac{1}{l}E^{A}\bm{Z}_{A},
\end{equation}
but this leads us nowhere; in order to describe gravity, it is imperative to include the generators $\bm{J}_{AB}$ of the Lorentz symmetry.
The set $\left\{\bm{Z}_{A}, \bm{J}_{AB}\right\}$, however, does not span a Lie algebra.
In order to close the algebra, the smallest possible modification is to add a single extra generator $\bm{D}$, leading us to the Lie algebra
\begin{align}
 \left[\bm{Z}_{A},\bm{Z}_{B}\right]&=l\left(\bm{Z}_{A}k_{B}-\bm{Z}_{B}k_{A}\right),\label{Eq_Weyl_1}\\
 \left[\bm{J}_{AB},\bm{Z}_{C}\right]&=\eta_{BC}\bm{Z}_{A}-\eta_{AC}\bm{Z}_{B}+ \nonumber \\
 & + l\left(k_{A}\eta_{BC}-k_{B}\eta_{AC}\right)\bm{D},\label{Eq_Weyl_2}\\
 \left[\bm{J}_{AB},\bm{J}_{CD}\right]&=\eta_{BC}\bm{J}_{AD}-\eta_{BD}\bm{J}_{AC}+ \nonumber \\
 & -\eta_{AC}\bm{J}_{BD}+\eta_{AD}\bm{J}_{BC},\label{Eq_Weyl_3}\\
 \left[\bm{D},\bm{Z}_{A}\right]&=\bm{Z}_{A}+lk_{A}\bm{D}.\label{Eq_Weyl_4}
\end{align}

This algebra is not as unusual as it may seem at first sight.
It suffices to make the change of basis
\begin{equation}
 \bm{P}_{A}=\bm{Z}_{A}+lk_{A}\bm{D}
\end{equation}
to see that it is equivalent to
\begin{align}
 \left[\bm{P}_{A},\bm{P}_{B}\right]&=0,
 \label{eq:Weylfirst} \\
 \left[\bm{J}_{AB},\bm{P}_{C}\right]&=\eta_{CB}\bm{P}_{A}-\eta_{CA}\bm{P}_{B},\\
 \left[\bm{J}_{AB},\bm{J}_{CD}\right]&=\eta_{BC}\bm{J}_{AD}-\eta_{BD}\bm{J}_{AC}+ \nonumber \\ 
 & -\eta_{AC}\bm{J}_{BD}+\eta_{AD}\bm{J}_{BC},\\
 \left[\bm{D},\bm{P}_{A}\right]&=\bm{P}_{A},
 \label{eq:Weyllast}
\end{align}
which is the Weyl subalgebra of the conformal algebra,
\begin{align}
 \left[\bm{P}_{A},\bm{P}_{B}\right]&=0,\\
 \left[\bm{K}_{A},\bm{K}_{B}\right]&=0,\\
 \left[\bm{K}_{A},\bm{P}_{B}\right]&=\eta_{AB}\bm{D}-\bm{J}_{AB},\\
 \left[\bm{J}_{AB},\bm{P}_{C}\right]&=\eta_{CB}\bm{P}_{A}-\eta_{CA}\bm{P}_{B},\\
 \left[\bm{J}_{AB},\bm{K}_{C}\right]&=\eta_{CB}\bm{K}_{A}-\eta_{CA}\bm{K}_{B},\\
 \left[\bm{J}_{AB},\bm{J}_{CD}\right]&=\eta_{BC}\bm{J}_{AD}-\eta_{BD}\bm{J}_{AC}+ \nonumber \\
 &-\eta_{AC}\bm{J}_{BD}+\eta_{AD}\bm{J}_{BC},\\
 \left[\bm{D},\bm{P}_{A}\right]&=\bm{P}_{A},\\
 \left[\bm{D},\bm{K}_{A}\right]&=-\bm{K}_{A}. 
\end{align}

Given this intimate relation between the Weyl subalgebra and the RS spacetime,
one may expect that the RS solution represents an interesting vacuum for a gauge theory invariant under this symmetry.
The Weyl CS gravity in $d=5$ is precisely this kind of gauge theory. 
The construction of the Lagrangian and the study of its dynamics are treated in the following sections.

\section{Chern--Simons gravity with Weyl invariance in five dimensions}
\label{sec:CStheory}

CS gravity theories have been studied with interest in the last decades.
The idea originated in the 1980s with the work of Refs.~\cite{Des82a,Des82b,Ach86,Ach89,Cha89,Cha90,Banh96a,Tro97},
and it has been developed in the context of gravity and supergravity since then. 
A recent review, with a focus on its application to gravity theory,
can be found in Ref.~\cite{Zan15}.

In odd dimensions, $d=2n+1$, a CS Lagrangian is defined as
\begin{equation}
 \mathcal{L}_{\mathrm{CS}}^{\left(2n+1\right)} =\kappa_{n}\int_{0}^{1}\mathrm{d}\tau\left\langle\bm{A}\wedge\left(\tau\mathrm{d}\bm{A}+\frac{1}{2}\tau^{2}\left[\bm{A},\bm{A}\right]\right)^{\wedge n}\right\rangle ,\label{Eq_CS_Lagrangian}
\end{equation}
where $\bm{A}$ corresponds to a Lie algebra-valued connection one-form
\begin{equation}
 \bm{A} = \nwse{A}{A}{\mu} \mathrm{d}x^{\mu} \otimes \bm{G}_{A},
\end{equation}
and
\begin{align}
 \left\langle\cdots\right\rangle :\mathfrak{g}^{n+1}&\rightarrow\mathbb{R}\\
 \left( \bm{G}_{A_{1}}, \ldots, \bm{G}_{A_{n+1}} \right) & \mapsto
 g_{A_{1} \cdots A_{n+1}} = \left\langle \bm{G}_{A_{1}} \cdots \bm{G}_{A_{n+1}} \right\rangle,
 \label{Eq_Invariant_Tensor}
\end{align}
stands for a multilinear symmetric form of rank $n+1$ invariant under the Lie algebra $\mathfrak{g}$.

This kind of Lagrangian has several attractive features as a field theory.
For our purposes, the most relevant are~\cite{Zan15}:
\begin{enumerate}
 \item CS gravities are background-free (a background vielbein is not necessary to construct the theory,
       as it is for instance in the case of the Yang--Mills Lagrangian).
 \item CS gravities are off-shell invariant under the symmetry group (up to boundary terms).
       This stands in strong contrast with the standard Einstein--Hilbert Lagrangian in $d=4$~\cite{Kib61,Mac77}.
 \item Despite its topological origin, this kind of Lagrangian has propagating degrees of freedom when $d\geq 5$
       (i.e., its degrees of freedom are not ``just topological'' for $d\geq 5$).
\end{enumerate}

It is compelling to observe that the Einstein--Hilbert Lagrangian with cosmological constant in $d=3$ corresponds to a CS three-form
and therefore the Lagrangian is off-shell gauge invariant, in stark contrast with the four-dimensional case.
It has been conjectured~\cite{Cha89,Zan12} that this invariance could be the underlying reason
for the renormalizability of three-dimensional gravity~\cite{Wit88,Wit89a,Wit89b}.
Trying to repeat this feat has been perhaps one the strongest motivations to pursue this kind of theory in higher dimensions.

The main challenge here lies in the dynamics. 
In $d=3$, the degrees of freedom are topological and not propagating;
the equations of motion are simply $\bm{F}=0$.
In $d\geq 5$, the equations of motion are highly nonlinear and the phase space has a complicated structure.
In fact, in general the standard Dirac method for constraints cannot be directly used~\cite{Saa01,Mis03,Mis05},
and this has been a big obstacle for the quantization of this kind of theory in higher dimensions.
It is interesting to notice that $\bm{F}=0$ is still a solution for the equations of motion in higher dimensions,
but it is somehow ``hyper-stable,'' meaning that perturbations on this background lead to the equation $0=0$ and do not propagate.
In practice, higher-dimensional CS theories can have solutions admitting propagating degrees of freedom, but finding these states is nontrivial.

In order to solve this problem, warped spacetime solutions have been successfully used
in the past as vacuum states in CS gravity theories in $d=11$ (see Refs.~\cite{Has03,Has05}).
In the present work, we will also use this kind of vacuum, but from a slightly different point of view:
the vanishing torsion condition is not imposed in the bulk, and the RS space corresponds to
the group manifold for the translational part of the algebra.
This means that on this vacuum the f\"{u}nfbein corresponds to the MC form for the translational piece. 

The theory is constructed using the Lagrangian~(\ref{Eq_CS_Lagrangian}) for $n=2$ (i.e., $d=5$)
when the one-form connection takes values in the Weyl subalgebra [cf.~eqs.~(\ref{Eq_Weyl_1})--(\ref{Eq_Weyl_4})]
\begin{equation}
 \bm{A}=\frac{1}{2}W^{AB}\bm{J}_{AB}+\frac{1}{l}E^{A}\bm{Z}_{A}+\phi\bm{D},\label{Eq_Weyl_Connection}
\end{equation}
where $E^{A}$ is the f\"{u}nfbein one-form,
$W^{AB}$ is the five-dimensional spin connection and
$\phi=\phi_{\mu}\mathrm{d}x^{\mu}$ is the one-form field associated to dilations.
The choice of the basis (\ref{Eq_Weyl_1})--(\ref{Eq_Weyl_4}) for the following work stems from practical
purposes, since aspects of the RS scenario are made explicit through the algebra and its commuting relations.

Under an infinitesimal gauge transformation generated by the local parameter
\begin{equation}
 \bm{\lambda}=\frac{1}{2}\lambda^{AB}\bm{J}_{AB}+\frac{1}{l}\lambda^{A}\bm{Z}_{A}+\lambda\bm{D},
\end{equation}
the fields transform as components of the connection one-form, i.e.,
\begin{equation*}
 \delta\bm{A}=-\left(\mathrm{d}\bm{\lambda}+\left[\bm{A},\bm{\lambda}\right]\right).
\end{equation*}

For Lorentz transformations we find
\begin{align}
  \delta W^{AB}& = -\mathrm{D}_{W}\lambda^{AB},\\
  \delta E^{A} & = \lambda^{A}{}_{B}E^{B},\\
  \delta \phi  & = k_{A}\lambda^{A}{}_{B}E^{B}\label{Eq_Var_Phi_Lorentz}.
 \end{align}
The $\bm{Z}_{A}$-boosts, on the other hand, act on the fields as
 \begin{align}
  \delta W^{AB} & =0,\\
  \delta E^{A}  & =-\mathrm{D}_{W}\lambda^{A}-k_{B}\lambda^{B}E^{A}+\left(E_{\bot}-\phi\right)\lambda^{A},\\
  \delta \phi   & =-k_{A}W^{A}{}_{B}\lambda^{B}-\phi k_{A}\lambda^{A}.
 \end{align}
Finally, dilations read
 \begin{align}
  \delta W^{AB} & = 0,\\
  \delta E^{A}  & = \lambda E^{A},\\
  \delta \phi   & = -\mathrm{d}\lambda+\lambda E_{\bot},
 \end{align}
where $\mathrm{D}_{W}$ stands for the usual Lorentz covariant derivative,
and $E_{\bot}=k_{A}E^{A}$.

It must be stressed that despite the nonstandard form of the commutator~(\ref{Eq_Weyl_2}),
the f\"{u}nfbein behaves as a Lorentz vector, and therefore the metric~(\ref{Eq_Metric})
remains invariant under local Lorentz transformations, as expected.
However, it is interesting to notice that, in our chosen basis [cf.~eqs.~(\ref{Eq_Weyl_1})--(\ref{Eq_Weyl_4})],
the dilation field $\phi$ is not a Lorentz scalar,
since Lorentz transformations on planes orthogonal to the brane are going to change it [cf.~eq.~(\ref{Eq_Var_Phi_Lorentz})].

The fundamental ingredient in order to shape the Lagrangian for the theory is the invariant tensor.
In order to construct a nontrivial invariant tensor, we will use a representation in terms of Dirac matrices
in $D=6$ for the generators of the algebra~(\ref{Eq_Weyl_1})--(\ref{Eq_Weyl_4}).
See Appendix~\ref{sec:Dirac} for further details.

Using this matrix representation, we find that the non-vanishing symmetric invariant tensor components read
\begin{align}
 \left\langle\bm{J}_{AB}\bm{J}_{CD}\bm{Z}_{E}\right \rangle&=\frac{1}{8\sqrt{2}}\epsilon_{ABCDE}+ \nonumber \\
 &+\frac{\alpha}{4}\left(\eta_{AC}\eta_{BD}-\eta_{AD}\eta_{BC}\right)lk_{E},\label{Eqs_Invariant_Tensor-ini}\\
 \left\langle\bm{J}_{AB}\bm{J}_{CD}\bm{D}\right\rangle&=-\frac{\alpha}{4}\left(\eta_{AC}\eta_{BD}-\eta_{AD}\eta_{BC}\right),\\
 \left\langle\bm{Z}_{A}\bm{Z}_{B}\bm{Z}_{C}\right\rangle&=-\left(\frac{3}{4}\alpha+\alpha^{3}\right)l^{3}k_{A}k_{B}k_{C},\\
 \left\langle\bm{Z}_{A}\bm{Z}_{B}\bm{D}\right\rangle&=\left(\frac{3}{4}\alpha+\alpha^{3}\right)l^{2}k_{A}k_{B},\\
 \left\langle\bm{Z}_{A}\bm{DD}\right\rangle&=-\left(\frac{3}{4}\alpha+\alpha^{3}\right)lk_{A},\\
 \left\langle\bm{DDD}\right\rangle&=\frac{3}{4}\alpha+\alpha^{3},\label{Eqs_Invariant_Tensor-fin}
\end{align}
where $\alpha$ is an arbitrary constant. 

Using the subspace separation method introduced in Refs.~\cite{Iza05,Iza06a},
it is possible to write down an explicit Lorentz-invariant expression for the CS Lagrangian in $d=5$ as
\begin{align}
 \mathcal{L}_{\mathrm{CS}}^{\left(5\right)}\left(\boldsymbol{A}\right)&=\frac{3}{32\sqrt{2}}\frac{1}{l}\epsilon_{ABCDE}R^{AB}\wedge R^{CD}\wedge E^{E}+\\
 &+\frac{3}{8}\alpha\left(\phi-E_{\bot}\right)\wedge R^{A}{}_{B}\wedge R^{B}{}_{A}+\\
 &+\left(\frac{3}{4}\alpha+\alpha^{3}\right)\left(\phi-E_{\bot}\right)\wedge\left(\mathrm{d}\phi-\mathrm{d}E_{\bot}\right)^{\wedge2},
\end{align}
where $R^{AB}=\mathrm{d}W^{AB}+W^{A}{}_{C}\wedge W^{CB}$ and $T^{A}=\mathrm{d}E^{A}+W^{A}{}_{B}\wedge E^{B}$
stand for the five-dimensional Lorentz curvature and torsion.

The field equations (obtained by regarding $E^{A}$, $W^{AB}$ and $\phi$ as independent fields)
can be reduced to the following independent relations in $d=5$:
\begin{align}
 \epsilon_{ABCDE}R^{AB}\wedge R^{CD}&=0\label{Eq_Var_E},\\
 \frac{1}{l}\epsilon_{ABCDE}R^{CD}\wedge T^{E}-\alpha R_{AB}\wedge \mathrm{d}\left(\phi-E_{\bot}\right)&=0,\label{Eq_Var_W}\\
 \alpha\left[\frac{1}{8}R^{A}{}_{B}\wedge R^{B}{}_{A}+\left(\frac{3}{4}+\alpha^{2}\right)\left[\mathrm{d}\left(\phi-E_{\bot}\right)\right]^{\wedge2}\right]&=0\label{Eq_Var_phi}.
\end{align}

Some comments on the field equations~(\ref{Eq_Var_E})--(\ref{Eq_Var_phi}) are in order.
In section~\ref{subsec:Torsional-RS-background}, a background solution with $R^{AB}=0$ and $T^{A}\neq 0$ is considered,
and in section~\ref{subsec:Perturbations}, the propagation of perturbations on this background is studied.

Of course, the ``hyper-stability'' problem
(also referred to as ``linearization instability'')
on the $R^{AB}=0$ background is still present in eq.~(\ref{Eq_Var_E}),
but \emph{not} in eq.~(\ref{Eq_Var_W}).
That is why the linearized dynamics of perturbations in section~\ref{subsec:Perturbations} stems from eqs.~(\ref{Eq_Var_W}) and~(\ref{Eq_Var_phi}).
Eq.~(\ref{Eq_Var_E}) cannot be simply dismissed, of course,
because in principle it could give rise to nonlinear constraints
which could seriously hinder the propagation of perturbations.
That this is not the case for the vacuum proposed by us
can be shown by considering the non-perturbative dynamics of the system.
This is done in section~\ref{subsec:Non-perturbative-dynamics},
where it is shown that in fact eq.~(\ref{Eq_Var_E}) does not prevent the propagation of perturbations on the torsional background.
However, eq.~(\ref{Eq_Var_E}) has nontrivial dynamical consequences:
the four-dimensional cosmological constant has to be replaced by a ``cosmological function''
related to the four-dimensional Euler density of the brane.

\section{Dynamics}
\label{sec:dynamics}



The goal of this section is to find a vacuum for the CS theory constructed in section~\ref{sec:CStheory}
that allows for linear perturbations of the fields to propagate, and to study its dynamics.
First, in section~\ref{subsec:Torsional-RS-background} a vacuum is constructed,
and propagation of linear perturbations on the field equations~(\ref{Eq_Var_W})
and~(\ref{Eq_Var_phi}) is studied in section~\ref{subsec:Perturbations}.
After this, the dynamics generated by the whole set of field equations,
including eq.~(\ref{Eq_Var_E}), is analyzed non-perturbatively in section~\ref{subsec:Non-perturbative-dynamics}.  

There are several possible vacuum candidates for the field equations~(\ref{Eq_Var_E})--(\ref{Eq_Var_phi}).
A warped space choice along the lines of Refs.~\cite{Has03,Has05} is a good solution to the problem.
Here we will use also a warped geometry solution, but with a twist: torsion doesn't have to vanish.
As we will show, in this context it is very natural to have torsion in the bulk, but a torsion-free brane universe.  
Here, torsion is regarded as a pure geometric quantity, along the lines of the Einstein--Cartan formalism.
It should be noted though, that torsion could emerge from other sources, as happens for instance when we add
fermions to a gravity theory in $d=4$, and in String Theory, where torsion arises from the Kalb--Ramond field potential.
When torsion is present, the spin connection and the vielbein represent independent degrees of freedom.
The full spin connection can be split as $W^{AB} = \mathring{W}^{AB} + \kappa^{AB}$,
where $\mathring{W}^{AB}$ stands for the torsion-free part, defined implicitly by
$\mathrm{d}E^{A}+\mathring{W}^{A}{}_{B}\wedge E^{B}=0$,
and $\kappa^{AB}$ is the cotorsion one-form tensor, which is related to the torsion two-form through
\begin{equation}
 T^{A}=\kappa^{A}{}_{B}\wedge E^{B}.
\end{equation}

In this case one must distinguish between the Lorentz curvature two-form,
$R^{AB}=\mathrm{d}W^{AB}+W^{A}{}_{C}\wedge W^{CB}$,
built from the full Lorentz connection $W^{AB}$,
and the Riemann curvature two-form,
$\mathring{R}^{AB}=\mathrm{d}\mathring{W}^{AB}+\mathring{W}^{A}{}_{C}\wedge \mathring{W}^{CB}$,
built from the torsion-free part only.
Both are related through the equation
\begin{equation}
 R^{AB}=\mathring{R}^{AB}+\mathring{\mathrm{D}} \kappa^{AB}+\kappa^{A}{}_{C}\wedge \kappa^{CB},
\end{equation}
where $\mathring{\mathrm{D}}$ stands for the covariant derivative in the torsion-free connection $\mathring{W}^{AB}$.

From now on, we will use the range of indices in uppercase as $A,B,C=0,\ldots ,4$ and $a,b,c=0,\ldots ,3$ in lowercase.

\subsection{Torsional Randall--Sundrum background solution}
\label{subsec:Torsional-RS-background}

Let us start by considering a RS geometry described by the f\"{u}nfbein
\begin{equation}
 E^{A}=\left\{
  \begin{array}
   [c]{l}%
   E^{a}=e^{-k\left \vert z\right \vert }m^{a},\label{Eq_Funfbein_RS_Minkowski}\\
   E^{4}=\mathrm{d}z,
  \end{array}
 \right.
\end{equation}
where $m^{a}$ corresponds to the vierbein of four-dimensional Minkowski spacetime.

The independent degrees of freedom of the spin connection described by the cotorsion can be chosen as
\begin{equation}
 \kappa^{AB}=-\mathrm{sgn}\left(z\right)\left(k^{A}E^{B}-k^{B}E^{A}\right).\label{Eq_Contorsion_Lorentz_Flat}
\end{equation}
This is a very particular state. 
Calling $\mathring{\omega}^{ab}$ the torsion-free spin connection for the brane
(defined by $\mathrm{d}m^{a}+\mathring{\omega}^{a}{}_{b}\wedge m^{b}=0$),
then the spin connection generated by the cotorsion~(\ref{Eq_Contorsion_Lorentz_Flat}) is given by
\begin{equation}
 W^{AB}=\left\{
 \begin{array}
 [c]{l}%
 W^{ab}=\mathring{\omega}^{ab},\\
 W^{a4}=0,
 \end{array}
 \right.
\end{equation}
It may seem paradoxical, but this connection creates a non-vanishing torsion in the bulk given by
\begin{equation}
T^{A}=\mathrm{d}E^{A}+W^{A}{}_{B}\wedge E^{B}=\mathrm{sgn}\left(  z\right)  E^{A}\wedge E_{\bot}.
\label{Eq_Torsion_Minkowski}
\end{equation}
This is a state with nonzero torsion in the bulk, but with vanishing torsion in the brane.
In a similar way, the standard Riemann curvature two-form corresponds as usual to
\begin{align}
\mathring{R}^{ab}  & =-\mathrm{sgn}^{2}\left(  z\right)  k^{2}E^{a}\wedge
E^{b},\\
\mathring{R}^{a4}  & =\left[  2\delta \left(  z\right)  -k~\mathrm{sgn}%
^{2}\left(  z\right)  \right]  E^{a}\wedge E_{\bot},
\end{align}
but the spacetime is Lorentz-flat, i.e., $R^{AB}=0$, and therefore the torsion is covariantly constant, $\mathrm{D}_{W}T^{A}=0$.%
\footnote{This is a consequence of the Bianchi identities, $\mathrm{D} R^{AB} = 0$, $\mathrm{D} T^{A} = \nwse{R}{A}{B} E^{B}$.}

Finally, the dilation one-form field is given by
\begin{equation}
 \phi=u\left(x\right)\mathrm{d}v\left(x\right),\label{Eq_Scalar_Flat}
\end{equation}
with $u\left(x\right)$ and $v\left(x\right)$ two arbitrary scalar functions depending on the brane coordinates.

\subsection{Perturbations}
\label{subsec:Perturbations}

The state described by eqs.~(\ref{Eq_Funfbein_RS_Minkowski}), (\ref{Eq_Contorsion_Lorentz_Flat}) and~(\ref{Eq_Scalar_Flat})
satisfies the field equations~(\ref{Eq_Var_E})--(\ref{Eq_Var_phi}) and proves to be an interesting vacuum for the CS theory.
As a matter of fact, let us consider a perturbation on the solution parameterized by
\begin{align}
 E^{A}       & \rightarrow E^{A}+\frac{1}{2}h^{A},\\
 W^{AB} & \rightarrow W^{AB}+U^{AB}+\Xi^{AB},\\
 \phi        & \rightarrow \phi+\varphi.
\end{align}

The perturbations of the spin connection are split in the mode $U^{AB}\left(h,\partial h\right)$,
which doesn't change the torsion, and the mode $\Xi^{AB}$, which does. In particular, this means that
\begin{equation}
 \frac{1}{2}\mathrm{D}_{W}h^{A}+U^{A}{}_{B}\wedge E^{B}=0,
\end{equation}
and $\delta T^{A}=\Xi^{A}{}_{B}\wedge E^{B}$.

The vacuum described by eqs.~(\ref{Eq_Funfbein_RS_Minkowski}), (\ref{Eq_Contorsion_Lorentz_Flat}),
and~(\ref{Eq_Scalar_Flat}) shows some interesting features.
First of all, since $R^{AB}=0$, eq.~(\ref{Eq_Var_E}) doesn't provide us with any information at the linear level
(the ``$0=0$'' problem), and therefore it will be studied non-perturbatively in section~\ref{subsec:Non-perturbative-dynamics}.
In contrast, eq.~(\ref{Eq_Var_W}) does allow for linear perturbations to propagate on this background.
A particularly interesting case from a four-dimensional point of view occurs when the one-form $h^{A}$ is given by the warped ansatz
\begin{equation}
 h^{A}=\left \{
  \begin{array}
   [c]{l}
   h^{4}=0,\\
   h^{a}=e^{-k\left \vert z\right \vert }\check{h}^{a}{}_{b}\left(x\right)e^{b},
  \end{array}
 \right.
\end{equation}
with $\mathrm{d}\phi=0$ and $\Xi^{AB}=0$.
In this case, linear perturbations on the field equation~(\ref{Eq_Var_W}) take the form
\begin{equation}
 \frac{1}{2}\epsilon_{abcd}\mathrm{D}_{\mathring{\omega}}U^{ab}\wedge e^{c}\wedge E_{\bot} =0,\label{Eq_Perturb-1}
\end{equation}
which correspond to standard gravitational waves
$\check{h}^{+}_{\mu\nu}=\frac{1}{2}\left(h_{\mu\nu}+h_{\nu\mu}\right)$
on a four-dimensional Minkowski background.
The general modes with $\Xi^{AB}\neq 0$ describe a ``torsional wave,'' 
which also depends on the antisymmetric component
$\check{h}^{-}_{\mu\nu}=\frac{1}{2}\left(h_{\mu\nu}-h_{\nu\mu}\right)$ of the one-form $h^{A}$.

This nontrivial dynamics for the propagation of linear perturbations seems highly satisfactory from a physical point of view.
However, its consistency with eq.~(\ref{Eq_Var_E}) at the non-perturbative level still has to be checked;
see section~\ref{subsec:Non-perturbative-dynamics}.

A more rigorous analysis of the symplectic matrix of the system,
\begin{equation}
 \Omega^{ij}{}_{KL}=-3g_{KLM}\epsilon^{ijkl}\bar{F}^{M}{}_{kl},
\end{equation}
following the lines of Ref.~\cite{Mis05},
but considering dilations instead of $U\left(1\right)$,
has been performed in the particular case when $\phi=\phi_{4}\left(x_{\mathrm{brane}}\right)E^{4}$.
The maximal rank of the matrix is $64-4=60$, but for the vacuum state described above the rank is just 32.
This is not a surprise considering that we are studying a vacuum state which is Lorentz flat, $R^{AB}=0$,
but it does indicate that we have an irregular and degenerate state, and the counting of degrees of freedom
cannot be performed naïvely following the standard Dirac procedure.

\subsection{Non perturbative dynamics}
\label{subsec:Non-perturbative-dynamics}

Let us slightly modify our ansatz.
First, let us consider a RS brane of arbitrary geometry, described by the f\"{u}nfbein one-form
\begin{equation}
 E^{A}=\left\{
  \begin{array}
   [c]{l}
   E^{a}=e^{-k\left\vert z\right\vert}e^{a},\\
   E^{4}=\mathrm{d}z,
  \end{array}
 \right.\label{Eq_General_Funfbein}%
\end{equation}
where now $e^{a}\left(x\right)$ is an arbitrary vierbein one-form for the brane
instead of the Minkowskian vierbein one-form $m^{a}$ we have used before.

In the same way, instead of eqs.~(\ref{Eq_Contorsion_Lorentz_Flat}) and~(\ref{Eq_Scalar_Flat})
for the cotorsion $\kappa^{AB}$ and the dilation one-form $\phi$, let us parametrize the fields as
\begin{align}
 \kappa^{AB} &=\left[\tau\left(x\right)e^{k\left\vert z\right\vert}-\mathrm{sgn}\left(z\right)\right]\left(k^{A}E^{B}-k^{B}E^{A}\right),\label{Eq_General_Contorsion}\\
 \phi        &=-2\left[kz\tau\left(x\right)+e^{-k\left\vert z\right\vert}\right]\mathrm{d}v\left(x\right),\label{Eq_General_Dilation}
\end{align}
where $\tau\left(x\right)$ and $v\left(x\right)$ are arbitrary scalar functions.

Under this ansatz, the components of the two-forms Lorentz curvature, torsion and dilation field strength read
\begin{align}
 R^{ab}         &=\mathring{r}^{ab}-k^{2}\tau^{2}e^{a}\wedge e^{b},\\
 R^{a4}         &=-k\mathrm{d}\tau\wedge e^{a},\\
 T^{A}          &=\left[\mathrm{sgn}\left(z\right)-\tau e^{k\left\vert z\right\vert}\right]E^{A}\wedge E_{\bot},\\
 \mathrm{d}\phi &=-2\left[\mathrm{sgn}\left(z\right)-\tau e^{k\left\vert z\right\vert}\right]e^{-k\left\vert z\right\vert}\mathrm{d}v\wedge E_{\bot} + \nonumber \\ & + 2kz\mathrm{d}v\wedge\mathrm{d}\tau,
\end{align}
where $\mathring{r}^{ab}=\mathrm{d}\mathring{\omega}^{ab}+\mathring{\omega}^{a}{}_{c}\wedge \mathring{\omega}^{cb}$
corresponds to the Riemann two-form associated to the torsionless spin connection $\mathring{\omega}^{ab}$ on the brane
(defined by $\mathrm{d}e^{a}+\mathring{\omega}^{a}{}_{b}\wedge e^{b}=0$).

Again, this is a torsionless configuration for the brane, but with non-vanishing torsion for the bulk.
Using the Bianchi identities it is straightforward to prove that
\begin{equation}
 \mathrm{D}_{W}T^{A}=-e^{k\left\vert z\right\vert}\mathrm{d}\tau\wedge E^{A}\wedge E_{\bot},
\end{equation}
and therefore $\tau=\text{const.}$ corresponds to a covariantly constant torsion state, $\mathrm{D}_{W}T^{a}=0$.

When the new ans\"{a}tze for the fields, eqs.~(\ref{Eq_General_Funfbein})--(\ref{Eq_General_Dilation}), are satisfied,
it is possible to prove that the field equations~(\ref{Eq_Var_E})--(\ref{Eq_Var_phi}) reduce to the independent relations for the four-dimensional geometry
\begin{align}
 \epsilon_{abcd}\mathring{r}^{ab}\wedge\mathring{r}^{cd}-k^{4}\tau^{4}\epsilon_{abcd}e^{a}\wedge e^{b}\wedge e^{c}\wedge e^{d}                           &=0,\label{Eq_Mov_Cosmo_Function}\\
 \frac{1}{2}\epsilon_{abcd}\left(\mathring{r}^{ab}-k^{2}\tau^{2}e^{a}\wedge e^{b}\right)\wedge e^{c}-\alpha kle_{d}\wedge\mathrm{d}v\wedge\mathrm{d}\tau &=0,\label{Eq_Mov_EH+Weird}\\
 \alpha \mathring{r}^{a}{}_{b}\wedge\mathring{r}^{b}{}_{a}                                                                                     &=0.\label{Eq_Mov_Pontryagin}%
\end{align}
Eq.~(\ref{Eq_Mov_Pontryagin}) implies that the four-dimensional brane geometry is restricted to be Pontryagin-density-vanishing when $\alpha \neq0$. Eq.~(\ref{Eq_Mov_EH+Weird}) corresponds to the standard four-dimensional Einstein--Hilbert equations,
with an effective four-dimensional stress-energy tensor given by
\footnote{Here $\ast$ denotes the four-dimensional Hodge dual.}
\begin{equation}
 \kappa_{\mathrm{4}}\mathcal{T}_{ab}\ast e^{a}=\alpha kle_{b}\wedge\mathrm{d}v\wedge\mathrm{d}\tau \label{Eff_SE_Tensor}
\end{equation}
and a ``cosmological function''
\begin{equation}
 \Lambda\left(x\right)=3k^{2}\tau^{2}\left(x\right).
\end{equation}
Finally, eq.~(\ref{Eq_Mov_Cosmo_Function}) indicates that this cosmological function $\Lambda\left(x\right)$ is not arbitrary,
but related to the brane's Euler density,
\begin{equation}
 \Lambda\left(x\right)=\sqrt{\ast\left(-\frac{9}{4!}\epsilon_{abcd}\mathring{r}^{ab}\wedge\mathring{r}^{cd}\right)}.
 \label{Eq_Cosmological_Function}
\end{equation}

Spherically symmetric solutions to this modified theory of gravity include the Schwarzchild solution only in the far-field limit.
In order to avoid unphysical consequences, one might wish to turn to the supersymmetric extension of the theory.
In this case, some components of the one-form gravitini in the super-connection
can play the role of dark matter fields non-minimally coupled to the geometry. 
This may allow for a more realistic geometry solution, which will be treated elsewhere.

Using the non-perturbative equations~(\ref{Eq_Mov_Cosmo_Function})--(\ref{Eq_Mov_Pontryagin}),
it is possible to check the consistency of the perturbative analysis made in section~\ref{subsec:Perturbations}. 
In order to focus our attention on the gravitational piece,
we have chosen a non-perturbative solution which is much more general for the curvature and torsion.
However, we have chosen a more restrictive $\phi$ solution,
in such a way that it satisfies $\mathrm{d}\phi\wedge\mathrm{d}\phi=0$ and kills many terms orthogonal to the brane.

After some algebraic manipulation, the ``problematic'' field equation~(\ref{Eq_Var_E}) becomes eq.~(\ref{Eq_Mov_Cosmo_Function}).
The same happens with eqs.~(\ref{Eq_Var_W}) and~(\ref{Eq_Mov_EH+Weird}), and eqs.~(\ref{Eq_Var_phi}) and~(\ref{Eq_Mov_Pontryagin}), respectively.

Now it is possible to understand why the linear perturbations in section~\ref{subsec:Perturbations}
were allowed to propagate in the form of standard gravitational waves on a four-dimensional Minkowskian background.
The first term in eq.~(\ref{Eq_Mov_EH+Weird}) generates the standard four-dimensional torsionless Einstein--Hilbert dynamics.
Eq.~(\ref{Eq_Mov_Cosmo_Function}) does not freeze the dynamics, but it requires the ``cosmological function'' given by eq.~(\ref{Eq_Cosmological_Function}). The background considered in section~\ref{subsec:Perturbations} is equivalent to $\mathring{r}^{ab}=\tau=0$.
In such a case, the cosmological function $\Lambda=3k^{2}\tau^{2}$ vanishes and it doesn't change under linear perturbations.
Therefore, torsion-preserving perturbations on such a background must reproduce standard four-dimensional gravitational waves on the brane,
as was shown in eq.~(\ref{Eq_Perturb-1}).

\subsection{Cosmological Models}

So far, we have studied solutions for our $d=5$ CS gravity theory when torsion is nonzero along the bulk.
We have seen that torsion itself has an effect upon the Einstein equations for the brane
[i.e., that torsion plays a role in the cosmological function~(\ref{Eq_Cosmological_Function})],
so it would be interesting to explore whether it somehow impacts the dynamics of certain systems, e.g., cosmological models.

In order to probe the possible behavior when matter is coupled to the theory (e.g., in a supersymmetric extension),
let us consider the cosmological toy model generated when the cosmological constant is replaced by the
expression in eq.~(\ref{Eq_Cosmological_Function}).
To this end, we use the effective Einstein equations for the brane~(\ref{Eq_Mov_EH+Weird}) with a spatial-flat FLRW metric,
along with the effective four-dimensional stress-energy tensor~(\ref{Eff_SE_Tensor}), taken to be that of a perfect fluid,
suitable for a cosmological treatment.

With the previous considerations, the cosmological function~(\ref{Eq_Cosmological_Function}) becomes
\begin{equation}
 \Lambda\left(t\right)=\frac{3}{c^{2}}\frac{\dot{a}}{a}\sqrt{\frac{\ddot{a}}{a}}=\frac{3}{c^{2}}\sqrt{-q}H^{2},
 \label{Eq_Cosmo_Lambda}
\end{equation}
where $q=-\ddot{a}a/\dot{a}^{2}$ is the cosmological deceleration parameter.
From eq.~(\ref{Eq_Cosmo_Lambda}), we see that our model requires $q<0$ in order not to generate imaginary terms
in the field equations, and therefore it is forced to produce accelerating universes only.

The modified FLRW equations for the brane read
\begin{align}
 \left(\frac{\dot{a}}{a}\right)^{2}-\frac{\dot{a}}{a}\sqrt{\frac{\ddot{a}}{a}} 
 &=\frac{1}{3}c^{2}\kappa_{4}\rho,\label{Eq_FLRM_Modf-1}\\
 2\frac{\ddot{a}}{a}+\left(\frac{\dot{a}}{a}\right)^{2}-3\frac{\dot{a}}{a}\sqrt{\frac{\ddot{a}}{a}} 
 &=-c^{2}\kappa_{4}p.\label{Eq_FLRM_Modf-2}
\end{align}
We can see that eqs.~(\ref{Eq_FLRM_Modf-1})--(\ref{Eq_FLRM_Modf-2}) together imply the energy balance equation
\begin{equation}
 \dot{\rho}+3\frac{\dot{a}}{a}(\rho+p)=0.
\end{equation}

Now if we choose an equation of state in the form of a barotropic fluid $p=w\rho$,
plug it into the energy balance equation, and solve the resulting system together with eq.~(\ref{Eq_FLRM_Modf-1}),
we obtain several families of solutions.
One of them is a scale factor of the form
\begin{equation}
 a\left(t\right)=a_{0}\exp\left(c\sqrt{\frac{\bar{\Lambda}}{3}}t\right),
\end{equation}
where $a_{0}$ and $\bar{\Lambda}$ are constants.
This alternative leads to a standard cosmology with a fixed cosmological constant
$\Lambda=\bar{\Lambda}$ and exponential growth,
so we won't pursuit it any further.

More interesting solutions are of the form
\begin{equation}
 a \left( t \right) = a_{p} \left( \frac{t}{t_{p}} \right)^{\alpha_{w}},
 \label{Eq_Cosmo_t_power}
\end{equation}
and
\begin{equation}
 a \left( t \right) = a_{0} \left( 1 + \frac{H_{0}}{\alpha_{w}} t \right)^{\alpha_{w}},
 \label{Cosmo_(1-Ht)_power}
\end{equation}
where the $\alpha_{w}$ exponent is given by
\begin{equation}
 \alpha_{w} = -\frac{4}{9\left(w+1\right)\left(w-\frac{1}{3}\right)}.
\end{equation}

Let us briefly consider both scenarios.
When the $\alpha_{w}$ exponent is positive, the solution describes a universe where at $t=0$
we have $a\left(0\right)=0$ and a divergent Hubble parameter,
\begin{equation}
  H\left(t\right)=\frac{\alpha_{w}}{t}.
\end{equation}
The interesting point is that we always have a negative constant deceleration parameter,
\begin{equation}
  q=-\left[\frac{3}{2}\left(w+\frac{1}{3}\right)\right]^{2},
  \label{Eq_Deceleration_Postive}
\end{equation}
implying an accelerating universe regardless of the value of $w$.

\begin{figure}
  \centering
  \includegraphics{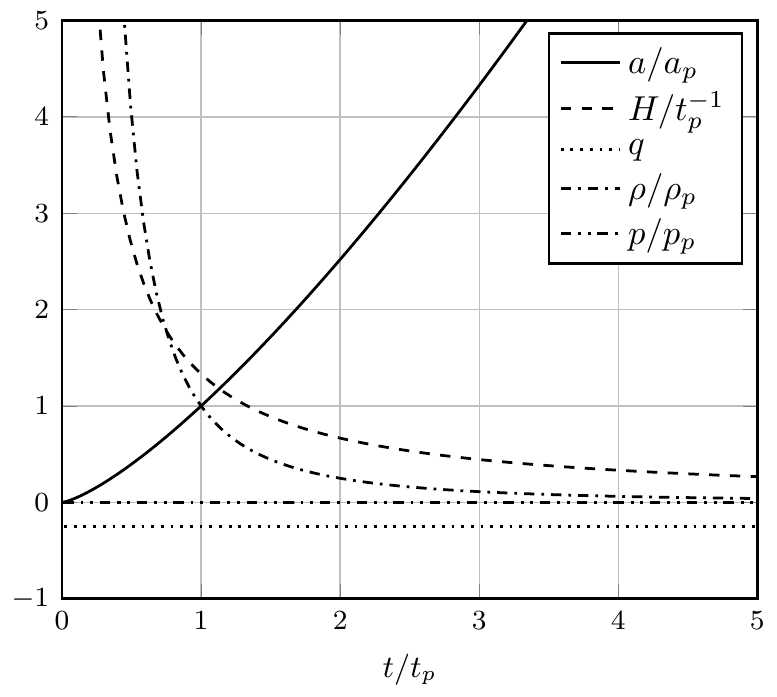}
  \caption{Cosmological parameters $a$, $H$, $q$, $\rho$ and $p$ for the
           solution in eq.~(\ref{Eq_Cosmo_t_power}), with $w=0$.
           The curve for $\Lambda$ is identical to the curve for $\rho$
           (properly normalized). When $w=0$, the pressure vanishes.}
  \label{Fig_Cosmo_00_a_H_q}
\end{figure}

Fig.~\ref{Fig_Cosmo_00_a_H_q}, drawn for the particular case of $w=0$ (dust), $\alpha_{w} = 4/3$,
is representative of the behavior of the cosmological parameters.

On the other hand, eq.~(\ref{Cosmo_(1-Ht)_power}) describes a scenario where at $t=0$
we have $a\left(0\right)=a_{0}$ and $H\left(0\right)=H_{0}$. 
Again, we find exactly the same constant and always negative deceleration parameter~(\ref{Eq_Deceleration_Postive}) from the former solution.
In this particular model, it is possible to find some singularities for $a\left(t\right)$ at finite future times depending in the values of $w$.
However, the important point is that the idea of a cosmological function given by eq.~(\ref{Eq_Cosmo_Lambda})
enforces a negative deceleration parameter.
 
It is worth observing that other ``self-accelerating'' models are known to arise from branes in higher-dimensional gravity (see Refs.~\cite{Def01,Def02,Dva07}).
Therefore, it seems natural to also find this kind of models in the context of the modified
FLRW equations~(\ref{Eq_FLRM_Modf-1})--(\ref{Eq_FLRM_Modf-2}) originated from a torsional RS state in five dimensional CS gravity.
However, the model shown here has the particular feature of always requiring acceleration in a natural way,
regardless of the details of the coupling to matter.
In particular, it is possible to have accelerated universes even in the case of only dust (dark matter).

\section{Conclusions}
\label{sec:final}

The main result of the present article is the realization that a flat RS brane is a solution of five-dimensional CS gravity,
and more importantly, it is a vacuum which admits the propagation of linear perturbations.
The CS theory is off-shell gauge invariant under the Weyl subgroup of $\mathrm{Conf}_{5}$.
Both, the solution and the symmetry, were carefully chosen in such a way that:
\begin{enumerate}
 \item The vacuum solution is a higher-dimensional generalization of a Bianchi space.
 \item The symmetry of the CS theory contains the Bianchi algebra as the translational subalgebra,
       and the f\"{u}nfbein is the piece of the connection associated to these generators.
\end{enumerate}
Together, both conditions imply that, on the vacuum, the f\"{u}nfbein
corresponds to the MC one-form for the translational part of the algebra.
In particular, this implies that the associated gauge curvature components vanish.

Our solution requires that five-dimensional torsion be nonzero.
It then becomes necessary to distinguish between Lorentzian and Riemannian curvature.
The chosen vacuum is indeed Lorentz-flat but not Riemann-flat. 
In this sense, it is necessary to follow an approach very similar to the one of Refs.~\cite{Alv14,Alv15}
but in five instead of three dimensions, and featuring a brane instead of a black hole.
And in the same way as in Refs.~\cite{Alv14,Alv15}, torsion is covariantly constant, $\mathrm{D}T^{A}=0$.

Despite all this, four-dimensional brane torsion vanishes
and the dynamics of the perturbations is the same as standard gravitational waves
in $d=4$ on a Minkowskian background.
However, far from the perturbative regime it induces an effective four-dimensional
``cosmological function'' $\Lambda\left(x\right)$ proportional to the square root of the Euler density of the brane,
\begin{equation}
 \Lambda\left(x\right)=\sqrt{\ast\left(-\frac{9}{4!}\epsilon_{abcd}\mathring{r}^{ab}\wedge\mathring{r}^{cd}\right)}
\end{equation}
This turns out to be an interesting toy model from the point of view of cosmology,
because the field equations only allow accelerated cosmologies with $q<0$,
regardless of the details of matter interaction. 
The effect of torsion in braneworld scenarios on the effective four-dimensional
Einstein equations has also been studied in Refs.~\cite{Hoff09a,Hoff09b,Khak09,Hoff09c,Hoff10}.

There is a lot of room for further research from the ideas presented here.

First, it is necessary to remember that what we have presented here is just a toy model,
and it would be interesting to construct a more realistic version of it.
Despite of having found a nice solution to the vacuum problem for this particular CS theory
and having some interesting possibilities in cosmology, it is necessary to stress
that having an effective ``cosmological function'' $\Lambda\left(x\right)$ proportional to
the square root of the Euler density also induces some unphysical consequences
(e.g., Schwarzschild solution only in the far-field limit).
However, the problem seems to be solvable by considering the supersymmetric extension
of the CS theory here presented (i.e., CS supergravity, see Refs.~\cite{Banh96a,Tro97}).
In this case, it is necessary to work with a super-connection one-form,
where the one-form gravitini fields are associated to the fermionic piece of the superalgebra.
This would relax some of the constraints, and the gravitini should play the role of 
a dark matter background for the brane geometry.
On this background, a more realistic dynamics is to be expected.
A somewhat similar system was studied in Ref.~\cite{Gar07}, where, however,
the CS theory is based upon a supersymmetric version of the anti-de~Sitter
algebra, instead of the Weyl algebra.

Second, it is possible to conjecture that the underlying reason for the nice behavior
of our vacuum choice is that, on it, spacetime has the structure of a group manifold,
with the f\"{u}nfbein corresponding to its associated MC one-form.
Here we have studied just one single case (the generalized Type V space of the Bianchi classification),
but many other possibilities remain to be studied.

\begin{acknowledgments}
The authors wish to thank Jorge Zanelli and Patricio Salgado for enlightening discussions,
as well as the two anonymous referees for their careful reading of the paper and their valuable suggestions.
FI is grateful to José~A. de~Azcárraga for his valuable suggestions regarding the treatment of Lie algebras.
This research was partially funded by Fondecyt grants 1130653 and 1150719 (FI),
and 3130444 (PM), and by Conicyt scholarship 22131299 (FC-T), from the Government of Chile.
\end{acknowledgments}

\appendix

\section{Representation of the Weyl Subalgebra}
\label{sec:Dirac}

In order to find a useful invariant tensor
[cf.~eqs.~(\ref{Eqs_Invariant_Tensor-ini})--(\ref{Eqs_Invariant_Tensor-fin})]
for the Weyl subalgebra~(\ref{Eq_Weyl_1})--(\ref{Eq_Weyl_4}),
it is necessary to find a suitable matrix representation.

Such a representation can be constructed using the Dirac representation in $D=6$,
\begin{equation}
 \Gamma_{A}\Gamma_{B}+\Gamma_{B}\Gamma_{A}=2\eta_{AB}\um,
\end{equation}
in terms of $2^{3} \times 2^{3}$ Dirac matrices.
Here we use uppercase indices with the range $A,B,C=0,1,\ldots,5$ and lowercase indices with the range $a,b,c=0,1,\ldots,4$.

In terms of the Dirac representation, it is possible to construct a base $\left \{\Gamma_{A_{1}\cdots A_{n}}\right \}_{n=0}^{6}$ for all $2^{3} \times 2^{3}$ matrices as
\begin{equation*}
 \Gamma_{A_{1}\cdots A_{n}}=\frac{1}{n!}\delta_{A_{1}\cdots A_{n}}^{B_{1}\cdots B_{n}}\Gamma_{B_{1}}\cdots \Gamma_{B_{n}},
\end{equation*}
where, with the exception of the identity matrix $\Gamma=\um$,
all $\Gamma_{A_{1}\cdots A_{n}}$ matrices are traceless.

In particular, let us consider the matrices
\begin{align*}
 \Gamma_{AB}&=\left\{
  \begin{array}
   [c]{c}
   \Gamma_{ab},\\
   \Gamma_{a5},
  \end{array}
 \right.\\
 \Gamma_{A}&=\left\{
 \begin{array}
  [c]{c}
  \Gamma_{a},\\
  \Gamma_{5}.
 \end{array}
 \right.
\end{align*}

In terms of them, a representation for the algebra~(\ref{Eq_Weyl_1})--(\ref{Eq_Weyl_4}) is given by
\begin{align*}
 \bm{J}_{ab}&=\frac{1}{2}\Gamma_{ab},\\
 \bm{Z}_{a} &=\frac{1}{2\sqrt{2}}\left(\Gamma_{a5}-\Gamma_{a}\right)-lk_{a}\left(\frac{1}{2}\Gamma_{5}+\alpha\um\right),\\
 \bm{D}     &=\frac{1}{2}\Gamma_{5}+\alpha \um,
\end{align*}
with $\alpha$ an arbitrary constant.

In order to construct the invariant tensor~(\ref{Eqs_Invariant_Tensor-ini})--(\ref{Eqs_Invariant_Tensor-fin}),
it is necessary to use the symmetrized trace, some properties of Dirac matrices in $D=6$, and the algebra of matrices
\begin{align*}
 \Gamma_{M_{1}\cdots M_{p}}\Gamma_{N_{1}\cdots N_{q}}&=\sum_{s=0}^{\min\left(p,q\right)}\frac{\left(-1\right)^{s\left(s-1\right)/2}}{\left(s!\right)^{2}\left(p-s\right)!\left(q-s\right)!}\times\\
                                                     &\times\eta_{\left[B_{1}\cdots B_{s}\right]\left[C_{1}\cdots C_{s}\right]}
                                                     \delta_{M_{1}\cdots\phantom{A_{p-s}B_{1}}\cdots M_{p}}^{A_{1}\cdots A_{p-s}B_{1}\cdots B_{s}}\times\\
                                                     &\times\delta_{N_{1}\cdots\phantom{C_{s}A_{p-s+1}}\cdots N_{q}}^{C_{1}\cdots C_{s}A_{p-s+1}\cdots A_{p+q-2s}}\Gamma_{A_{1}\cdots A_{p+q-2s}},
\end{align*}
where
\begin{equation*}
 \eta_{\left[  A_{1}\cdots A_{s}\right]  \left[  B_{1}\cdots B_{s}\right]}=\eta_{A_{1}C_{1}}\cdots \eta_{A_{s}C_{s}}\delta_{B_{1}\cdots B_{s}}^{C_{1}\cdots C_{s}}.
\end{equation*}

For more details on higher-dimensional Dirac matrices manipulation, a valuable guide can be found in Ref.~\cite{VanPro99}.

\bibliographystyle{utphys} 
\bibliography{biblio2016}

\end{document}